\documentclass[aps,prl,twocolumn,groupedaddress,showpacs]{revtex4}

\usepackage{graphicx}

\begin{document}

\title{Matter-wave Interferometry with Phase Fluctuating Bose-Einstein Condensates}

\author{G.-B. Jo, J.-H. Choi, C.A. Christensen, Y.-R. Lee, T.A. Pasquini,
  W. Ketterle, and D.E. Pritchard}

\homepage[URL: ]{http://cua.mit.edu/ketterle_group/}

\affiliation{MIT-Harvard Center for Ultracold Atoms, Research
Laboratory of Electronics, Department of Physics, Massachusetts
Institute of Technology, Cambridge, MA 02139, USA}

\date{\today}

\begin{abstract}
Elongated Bose-Einstein condensates (BECs) exhibit strong spatial
phase fluctuations even well below the BEC transition temperature.
We demonstrate that atom interferometers using such condensates are
robust against phase fluctuations, i.e. the relative phase of the
split condensate is reproducible despite axial phase fluctuations.
However, larger phase fluctuations limit the coherence time,
especially in the presence of some asymmetries in the two wells of
the interferometer.
\end{abstract}


\maketitle

A non-interacting zero temperature Bose-Einstein condensate is the
matter-wave analogue to the optical laser, and therefore the ideal
atom source for atom interfero\-metry.  Finite temperature and
atomic interactions profoundly change the coherence properties of a
condensate and introduce phase fluctuations and phase diffusion.
Those phenomena are of fundamental
interest~\cite{PHV00,PSW00,PSW01,DHR01,LYQ96,CDR97}, but also of
practical importance because they may limit the performance of atom
interferometers~\cite{SSP04,SHA05,HLF07}.  This applies in
particular to magnetic microtraps and waveguides (e.g. atom
chips)~\cite{FZM07} since tight confinement and elongated geometry
enhances phase diffusion and phase fluctuations.

Phase \emph{diffusion} is a quantum effect associated with the
coherent splitting of the condensate.  Number fluctuations lead to
density fluctuations, which, due to interactions, cause
fluctuations of the energy and cause diffusion of the relative
phase proportional to the chemical potential times $\Delta N/N$,
the amount of fluctuations in the relative atom number. In our
previous work~\cite{JSW07,JCC07}, we showed that such phase
diffusion could be dramatically reduced by number squeezing,
increasing the coherence time.  In this paper, we characterize and
discuss the role of spatial phase \emph{fluctuations} in an atom
interferometer.

Phase \emph{fluctuations} cause the condensate to break up into
several quasi-condensates with random phase, i.e. long range
coherence is lost.  This usually happens in elongated geometries
when the temperature is sufficiently high to excite such
modes~\cite{PSW01,DHR01}, or in interacting one-dimensional
condensates  even at zero temperature due to quantum fluctuations
~\cite{BAI06}. Spatial phase fluctuations have two major
consequences for atom interferometry. First, they speed up phase
diffusion, since $\Delta N/N$ refers now to the atom number in a
single quasi-condensate. Second, they make the atom interferometer
much more sensitive to random relative displacements of the split
condensates, which have to be smaller than the coherence length,
which, for condensates with phase fluctuations, can be much smaller
than the size of the condensate.

A typical elongated trap geometry, realized by an atom chip, has an
aspect ratio of $\sim$200~\cite{SJP05,SHA05,JSW07}, sufficient to
induce phase fluctuations in a quasi-condensate along the axial
direction~\cite{PSW01} already at very low temperatures (or in the
1D case, even at zero temperature). When the temperature of a
condensate is above the characteristic temperature, $T^* = {15 N
(\hbar \omega_z)^2}/{32 \mu }$ where $\mu$ is the chemical
potential, $N$ total atom number, $\omega_z$ axial trap frequency,
and $\hbar$ the Planck's constant divided by $2\pi$~\cite{PSW01},
then thermal excitations of low energy axial modes lead to
longitudinal phase fluctuations.  For temperatures above $T^*$, the
coherence length $L^*$ of a phase-fluctuating condensate is shorter
than the length $L$ of the condensate $L^*/L =T^*/T$~\cite{PSW01}.

Previous experiments~\cite{SJP05,SHA05,JSW07} on atom interferometry
have operated in a regime, where phase fluctuations are predicted to
be present. However, their presence has not been observed because
the interferometer was read out by integrating the interference
fringes along the axial direction.

In this paper we observe the axial phase fluctuations spatially
resolved and characterize their effect on the coherence time of
the atom interferometer.  We show explicitly, that atom
interferometry can be performed in the presence of phase
fluctuations.  This has been expected ~\cite{BAI06}, since for
sufficiently short times after splitting, those fluctuations are
identical for both condensates and therefore don't affect the
measurement of the relative phase. However, already at short
times, they degrade the contrast and can limit the coherence time.
As we discuss below, we believe that this degradation is not due
to the quantum effect of the increased relative number
fluctuations in each quasi-condensate because of the high degree
of number squeezing, but is rather caused by asymmetries in the
double well potential leading to relative motion of the
condensates.

\begin{figure}
\begin{center}
\includegraphics[width=3.2in]{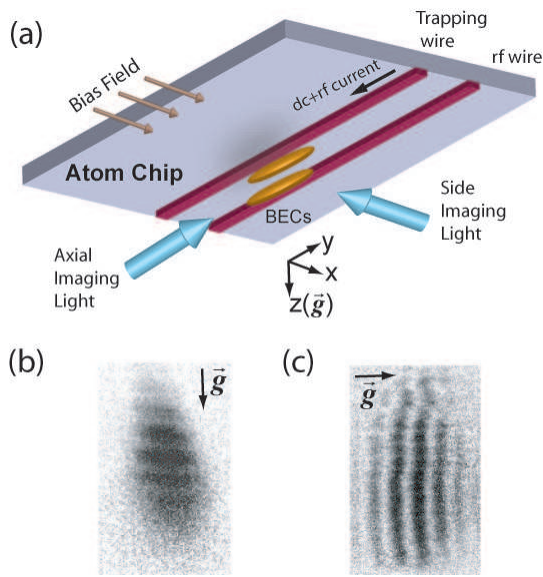}
\caption{(Color online) Geometry of the atom chip interferometer.
(a) Atoms were confined radially by the combined magnetic potential
of a current-carrying wire and an external bias field. A pair of
endcap wires (not shown) provided  axial confinement. The single
well was deformed into a vertical double well within 15~ms by adding
rf current into the trapping wire dressing the atoms with
oscillating rf fields.  Absorption image was taken by a probe beam
directed along the condensate axis [(b), axial imaging] and
perpendicular to the condensate axis [(c), side imaging]. All data
in this paper were obtained using  side imaging. The fields of view
are 160 $\times$ 260 $\mu$m and 180 $\times$ 100 $\mu$m for axial
and side imaging respectively.}
\end{center}
\end{figure}

Bose-Einstein condensates of $\sim4\times10^{5}$ $^{23}$Na atoms
in the $|F=1, m_{F}=-1\rangle$ state were transferred into a
magnetic trap generated by the trapping wire on an atom chip and
external bias field~\cite{SJP05}. Using adiabatic rf-induced
splitting~\cite{ZGT01,SHA05}, a double-well potential in the
vertical plane (parallel to the gravity direction) was formed as
illustrated in Fig. 1(a). Gravity was compensated by a magnetic
field gradient from the trapping wire. Typically, the separation
of the two wells was $d \sim 6~\mu$m, the height of the trap
barrier was $U \sim h \times 10$~kHz, and the difference of the
trap bottom between two wells $\sim h \times 300$~Hz. The trapping
frequencies were $\sim$~2~kHz (radial) and $\sim$~10~Hz (axial).
The absorption imaging light for data acquisition was resonant
with the $|F=2\rangle\rightarrow |F'=3\rangle$ cycling transition
for the trapped atoms and was aligned perpendicular to the
condensate axis [side imaging in Fig. 1(c)]. The atoms were
optically pumped into the $|F=2\rangle$ hyperfine level with a
pulse resonant with the $|F=1\rangle\rightarrow |F'=2\rangle$
transition.

\begin{figure}
\begin{center}
\includegraphics[width=2.8in]{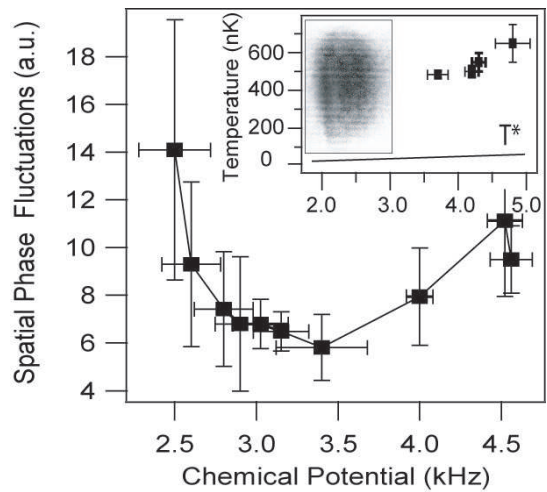}
\caption{ Spatial phase fluctuations in a single condensate.  The
phase fluctuations  were characterized by observing the density
modulations in an absorption image of the expanded cloud after 8~ms
time-of-flight (see inset) and calculating the rms fluctuations (see
text). The chemical potential (or atom number) was controlled by
additional rf-evaporative cooling. The temperature of the condensate
is shown in the inset graph. For chemical potentials less than
3.5~kHz, we could not measure the temperature of a condensate due to
the lack of discernable thermal atoms.  The observed phase
fluctuations do not decrease monotonically, but show a minimum at
the chemical potential of $\sim$3.5kHz, probably because the loss in
atom number compensated for the lower temperature . In the inset
graph, $T^*$ displays the characteristic temperature for the onset
of the phase fluctuations.}
\end{center}
\end{figure}

First, we characterized the presence of phase fluctuations in the
condensate before splitting by observing density modulations of the
expanded atomic cloud after 7~ms time-of-flight (Fig. 2 inset). In
trap, the mean-field interaction energy suppresses density
fluctuations, but ballistic expansion converts phase fluctuations
into density modulations~\cite{DHR01} since the initial velocity
field is proportional to the gradient of the phase. The number of
observed density striations of around ten is consistent with the
ratio of the measured temperature of $\sim$~650~$\pm$~100~nK and the
calculated value of $T^{*}\simeq 60~nK$.

The longitudinal phase fluctuations were quantified by measuring the
root-mean-square average of the density fluctuations as described in
Fig. 2~\cite{fluctuations}.  The amount of phase fluctuations was
controlled by changing the atom number and the temperature with
rf-evaporation. The rf field generated by the rf wire [Fig. 1(a)]
was swept down from $\sim10$~kHz above the Larmor frequency at the
trap center to a variable final value, leading to a variable
chemical potential and temperature of the condensate (Fig. 2 inset).
The variation of the spatial phase fluctuations with chemical
potential is shown in Fig. 2.

Having firmly established the presence of phase fluctuations, we can
now demonstrate the robustness of an atom interferometer against
longitudinal phase fluctuations.  For this, we split the condensates
and observe the reproducibility of interference fringes obtained by
recombining the condensates during ballistic expansion. The regular,
almost straight interference fringes (Figs. 1 and 3) show that the
spatial phase fluctuations are common mode and don't affect the
relative phase in a major way.

However, when we increase the amount of phase fluctuations, we
observe an increasing blurring or waviness of the interference
fringes (Fig. 3).  The number of wiggles of the waviness is
comparable to the modulation pattern observed in the ballistic
expansion of single condensates (Fig. 2).  For the smallest amount
of spatial phase fluctuations, the relative phase is almost constant
along the axial direction [dashed line in Fig. 3(c). The effect of
larger phase fluctuations is displayed by the solid line. However,
an average relative phase can still be determined.

To quantify the reproducibility of the relative phase, we determine
the probability of random phase (called randomness)~\cite{JSW07}
(Fig. 4).  For values of the chemical potential larger than 3.0~kHz,
the randomness is less than 0.1 which implies a reproducible phase
with 90$\%$ confinence. However, by comparing Figs. 2 and 4, one
clearly recognizes the degradation of  reproducibility of the
relative phase with increasing spatial phase fluctuations.

\begin{figure}
\begin{center}
\includegraphics[width=3.3in]{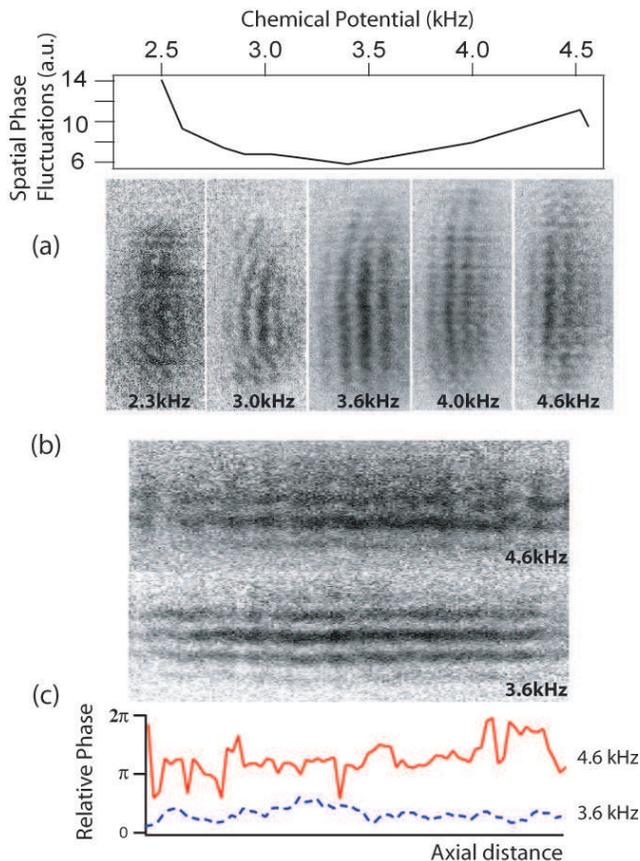}
\caption{ (Color online) Effect of spatial phase fluctuation on the
waviness of interference fringes. (a) Interference fringes obtained
right after splitting a condensate.  For large spatial phase
fluctuation (e.g. 4.6~kHz), the fringe pattern shows more
significant wiggles than for smaller phase fluctuations (e.g.
3.6~kHz). (b,c) From the fringes for 3.6~kHz (dashed line) and
4.6~kHz (solid line) chemical potentials, relative phases are
obtained along the axial direction. In both cases, the overall
relative phase can be well-determined by averaging along the axial
coordinate, but considerable axial variations of the relative phase
were observed in the regime of large longitudinal phase fluctuations
(solid line).}
\end{center}
\end{figure}

\begin{figure}
\includegraphics[width=2.8in]{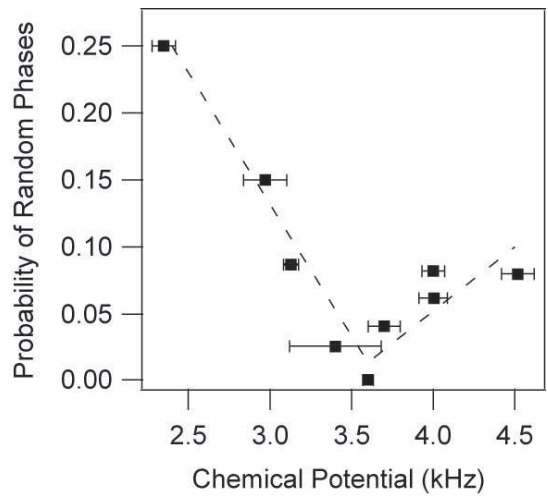}
\caption{ Effect of spatial phase fluctuations on the
reproducibility of the  relative phase right after splitting. The
probability of random phases was measured with variable longitudinal
phase fluctuations immediately after splitting (0~ms hold time). }
\end{figure}

By introducing a variable hold time after the splitting, we can
examine how spatial phase fluctuations limit the coherence time of a
matter-wave interferometer. Fig. 5 shows the increase of randomness
with hold time.  For the smallest amount of phase fluctuations
(chemical potential $\sim$3.4~kHz, black squares in Fig. 5), the
phase coherence time is $\sim$23~ms.  As the spatial phase
fluctuations increase (solid circles and open squares in Fig. 5),
the  phase coherence time becomes shorter.  It should be noted that
in the absence of spatial phase fluctuations, for a condensate with
zero temperature, the rate of phase diffusion decreases with
chemical potential, proportional to
$\sim\mu^{-1/4}$~\cite{LYQ96,CDR97}, which is also valid at finite
temperature~\cite{LYQ96}.  Our observed \emph{increase} of
decoherence with increasing chemical potential is therefore
attributed to the increase of spatial phase fluctuations.  The
increasing waviness of the interference fringes show that the
decoherence is caused by randomization of the relative phase along
the axial direction [Fig. 5(b)].

\begin{figure}
\begin{center}
\includegraphics[width=3.1in]{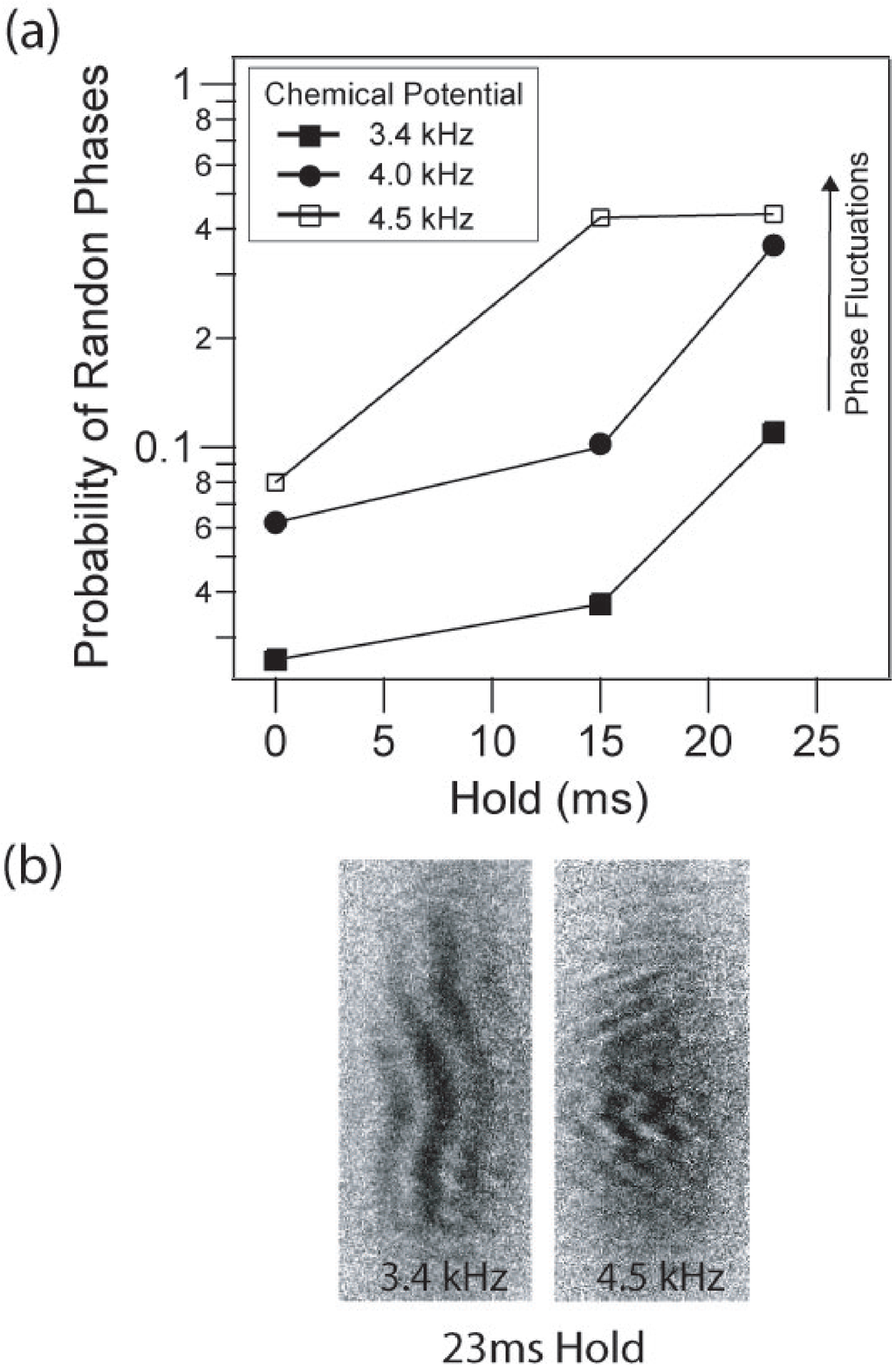}
\caption{ Effect of longitudinal phase fluctuations on the coherence
time between the split condensates.  (a) The probability for a
random phase for ten measurements of the relative phase is shown for
three different amounts of the longitudinal phase fluctuations. (b)
For condensates in the regime of large longitudinal phase
fluctuations ($\sim$4.5~kHz), interference fringes show more wavy
patterns, which led to the increased randomness of the measured
relative phase.}
\end{center}
\end{figure}

By which mechanism do the spatial phase fluctuations affect the
interferometer signal?  For our experimental parameters, the rate of
phase diffusion (assuming Poissonian number fluctuations after the
splitting) is $\sim$20~ms~\cite{LYQ96,CDR97}.  For our value of
$T/T^*$, the condensate fragments into $\sim$10 quasicondensates
which should decrease the coherence time by a factor of $\sqrt{10}$
to about 7 ms. Our observation of much longer coherence times
implies strong squeezing of the relative number fluctuations, as
already observed in Ref.~\cite{JSW07}.  In Ref.~\cite{JSW07} we
inferred a reduction of the number fluctuations below shot noise by
a factor of ten. However, having now established the presence of
strong phase fluctuations, we should reinterprete our previous
result. Those data were taken at a value of $T/T^*$ of about 7,
which implies that the number fluctuations for each quasi-condensate
was squeezed by a factor of $\sim$25.  Our current experiments were
carried out in a rotated geometry (in order to be able to observe
along a radial direction), but the value of $T/T^*$ $\sim$ 10 is
similar.  If we assume that the squeezing factor is the same, then
we should have observed phase coherence times comparable to the
200~ms observed previously~\cite{JSW07}.

We therefore conclude that the shorter coherence times observed in
this paper are not limited by the fundamental quantum phase
diffusion of quasi-condensates because of strong number squeezing,
but rather reflect the interplay of spatial phase fluctuations and
some random relative motion of the two condensates after splitting.
This is probably due to some asymmetries in the current trapping
potential~\cite{potential}.  The loss of coherence due to phase
fluctuations starts already during the splitting process (Figs. 3
and 4), and increases with hold time.

The main conclusions of this paper are that matter wave
interferometers are robust against spatial phase fluctuations,
especially when strong number squeezing mitigates the
fragmentation into smaller quasi-condensates (which show faster
phase diffusion than a single condensate).  However, spatial phase
fluctuations make the interferometer much more sensitive to
residual relative motion of the two split condensates and
therefore require a highly symmetric double well potential.

This work was funded by DARPA, NSF,and ONR.  G.-B. Jo and Y.-R. Lee
acknowledge additional support from the Samsung foundation. We thank
H.Kim for experimental assistance and Y.Shin for critical reading of
the manuscript. We also thank E. Demler for stimulating discussions.


\begin{thebibliography}{15}
\bibitem{fluctuations}
 For quantitative analysis of the spatial phase
fluctuations, the  density profile obtained from an absorption image
is Fourier transformed.  After filtering out high and low momentum
components, the profile is transformed back and its root-mean-square
average is determined (Fig. 2).
\bibitem{potential}
 In our trap geometry, the two condensates in
vertically separated potential wells feel different magnetic fields
from a pair of endcap wires. The slightly asymmetric axial
confinement leads to some quadrupolar relative motion of the two
separated condensates.  For the determination of the relative phase,
we select the central region where the fringes were parallel to the
axial direction.
\end{thebibliography}
\end{document}